\documentclass[reprint,aps,pra,amsmath,amsfonts]{revtex4-2}

\usepackage{amsmath}
\usepackage{amssymb}
\usepackage{graphicx}
\usepackage[colorlinks=true, allcolors=blue]{hyperref}
\usepackage{bm}
\usepackage{layout}
\usepackage{wrapfig}
\usepackage{paralist}
\usepackage{color}
\usepackage{titlesec}
\usepackage[utf8]{inputenc}
\usepackage[title]{appendix}

\begin{document}

\title[No Flow Noise]{A general approach to the statistics of microbial orientation: \\ L\'{e}vy walks, noise, and deterministic motion} 

 \author{Taylor J. Whitney$^a$}%
 \email{twhitney@ucmerced.edu}
\author{Thomas H. Solomon$^b$}
\email{tsolomon@bucknell.edu}
\author{Kevin A. Mitchell$^a$}
\email{kmitchell@ucmerced.edu}
\affiliation{ Univeristy of California: Merced$^a$ \text{5200 Lake Rd, Merced, CA 95343}, \text{and} Bucknell University$^b$ \text{One Dent Dr, Lewisburg, PA 17837}
}%

\date{\today}

\begin{abstract}
Microbial motion is typically analyzed by simplified models in which trajectories exhibit straight runs (perhaps with added Gaussian noise) followed by random, discrete tumbling events. We present the results of a statistical analysis of the angular dynamics for four different swimming microbes: tumbling and smooth-swimming strains of \textit{Bacillus subtilis} and two Eukaryotic algae, \textit{Tetraselmis suecica} and \textit{Euglena gracilis}. We show that the angular statistics closely resemble a Voigt profile, the convolution of a Gaussian (L\'{e}vy index $\alpha=2$) and Lorentzian (L\'{e}vy index $\alpha=1$) distribution. This distribution is ubiquitous for all four microbes. Rather than modeling tumbling  as a discrete process, we model tumbling dynamics as a continuous process: L\'{e}vy flights in the orientational dynamics using a Lorentzian noise model. This model is analytically solvable. Each individual microbe trajectory has both stochastic behavior (noise) and varying deterministic behavior, such as helices of different sizes and frequencies and circular arcs with different radii. We model the distribution of different deterministic behavior via an ensemble theory. The deterministic behavior (e.g., circular arcs) comes from physical observations of the swimming behavior and explains many of the qualitative features in the data that cannot be explained by a pure noise model. From this theory, we estimate the strength of Lorentzian noise, the physical rotational diffusion constant, and some relevant parameters relating to the distributions of deterministic behavior. This analysis shows that in some cases Gaussian noise is not the dominant process responsible for the angular statistics following a Voigt profile.
\end{abstract}

\maketitle

\section{\label{sec:Intro}Introduction}

Research interest in active matter systems has grown rapidly in recent years. Active matter covers a broad range of nonequilibrium physical systems. Some of the systems of great interest in this field include the behavior of self-propelled particles and active diffusion. Examples of these systems include swimming microbes \cite{koch11,Lushi14,Drescher11,Elgeti15} and self-propelled Janus particles \cite{Hu12,Gomez-Solano16}. 

L\'{e}vy walks have been used to model many phenomena in physical and biological systems, such as the search patterns of albatross, bumblebees, and deer \cite{Edwards07}, chaotic advection of passive tracer particles in flows~\cite{Solomon93}, \textit{E. coli} exploring its environment \cite{Huo21}, and human mobility patterns~\cite{Brockmann06}. There are also processes that are reminiscent of L\'{e}vy walks, such as bacterial hopping and trapping in pourous media \cite{Bhattacharjee19}. L\'{e}vy walks turn out to be a better search strategy in many scenarios than simple diffusion due to the super-diffusive behavior of the walks \cite{Viswanathan08}.

L\'{e}vy walks are trajectories with jumps defined by L\'{e}vy-$\alpha$ stable distributions, where the probability of the jump length $\ell$ is $P(\ell)\sim \ell^{-(\alpha+1)}$. L\'{e}vy walks are characterized by $0<\alpha<2$. These distributions have infinite variance for $\alpha<2$ and undefined mean for $\alpha \leq 1$. They are characterized by heavy tails, meaning they have a high probability of making arbitrarily large jumps. In contrast, if the L\'{e}vy index $\alpha=2$, we get a Gaussian distribution with well defined mean and variance. 

Microbial motion is often modeled via a run-and-tumble mechanism, with random discrete tumbling events interspersed between nearly straight runs.  We propose a different perspective on microbial dynamics by modeling the orientation via a continuous L\'{e}vy walk process. We test this model experimentally by monitoring the orientation of four different non-interacting microbial populations in two dimensions.  This model closely reproduces the experimental orientation statistics of these ensembles. In addition to stochastic processes, microbes exhibit a range of deterministic swimming behavior. Examples of these deterministic behaviors include helices~\cite{Constantino16} and circular arcs~\cite{Lauga06} (near solid surfaces). The variations in these deterministic behaviors affect the statistics in these systems. To our knowledge, the range of these behaviors has not been modeled in the literature. Therefore, in addition to the stochastic dynamics, we include such deterministic behavior as well.

\begin{figure*}
    \includegraphics[width=1.0\linewidth]{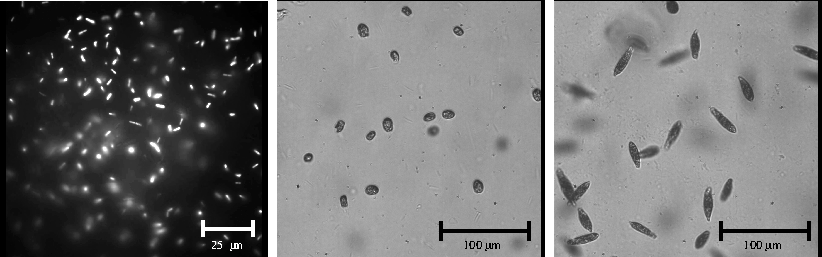}
    \caption{(a) \textit{B. subtilis} at 100X magnification. (b) \textit{Tetraselmis} at 40X magnification. (c) \textit{Euglena} at 40X magnification. }
    \label{fig:Microbes}
\end{figure*}

The paper is divided into several sections. Section~\ref{sec:Background} covers the background physics and current models used in the literature. Section~\ref{sec: Experiment} outlines the experiment and data collection for several different microbes.  Section \ref{sec: Stats} covers the statistical methods and applies them to the experimental data.  Section \ref{sec: Theory} outlines the proposed theory, starting from the stochastic differential equation (SDE) for the orientational dynamics. It then progresses into a Fokker-Planck equation for the system. Finally, it outlines the ensemble theory for randomly distributed drift parameters and addresses physically relevant applications of the theory.  Section \ref{sec: Results} discusses the application of this theory to the experimental data. Finally Sec.~\ref{sec: Conclusion} ends with discussion and closing remarks.

\section{Background of microbe swimming\label{sec:Background}}

At the length and velocity scales of many microorganisms, the Reynolds number is very small~\cite{Lauga20}. Thus, micro-swimmers move without relying on the inertia of the fluid. In low Reynolds number swimming, there is a balance between propulsion and drag, causing micro-swimmers whose mode of locomotion is essentially periodic to move at relatively constant speed~\cite{Lauga20}, though others may exhibit variations in speed. Though the mode of locomotion between prokaryotes and eukaryotes is fundamentally different, the fluid dynamics principles that allow them to move are the same: the breaking of time-reversal symmetry in the sequence of shape geometries~\cite{Purcell77}.

Prokaryotes rotate passive helical flagellar filaments using an intricate molecular machine called a bacterial rotary motor. The motor is driven via a proton pump mechanism~\cite{Lauga20}. Prokaryotes often have multiple flagella whose collective motion provides locomotion. The distribution of rotary motors can sometimes be asymmetric with the body of the bacterium. This leads to a range of trajectory types such as helices~\cite{Constantino16}. When the flagellar filaments unbundle, the microbes tumble~\cite{Elgeti15}. 

Eukaryotes on the other hand use a system more similar to muscle fibers. The internal structure of the flagella (called the axoneme) uses molecular machines (dynein) driven by ATP to slide passed each other causing the flagella to bend in an asymmetric periodic pattern~\cite{Lauga20}. During this process elastic forces build up, inevitably causing the microbe to stall and then tumble~\cite{Elgeti15}. 

A simplified model of this system suggests that one can model tumbling behavior as straight runs, followed by randomly distributed discrete tumbling events~\cite{Solon15}. In this model one considers a scattering function for the reorientation, which is distributed in time as a Poisson process, where tumbling events are exponentially distributed in time.  However, this does not capture all the random motion of the swimmers. The stochastic behavior during runs is often modeled as Gaussian noise and attributed to physical translational or rotational diffusion.

\section{\label{sec: Experiment}Experimental Data Collection}

The microbes studied in these experiments (Fig.~\ref{fig:Microbes}) are (a) {\em Bacillus subtilis}, small, prokaryotic microbes with a length that ranges approximately from 2 - 6 $\mu$m with width of approximately
0.5 - 1 $\mu m$; we study both a mutated, ``smooth-swimming'' strain (OI4139) and a ``tumbling'' strain (1A1266) with 
the GFP (green fluorescent protein) gene; (b) {\em Tetraselmis suecica}, marine (salt-water), eukaryotic
algae that are almost circular with typical diameter 10 - 15 $\mu$m; and (c) {\em Euglena gracilis}, freshwater algae with
length 40 - 80 $\mu$m and width 8-12 $\mu$m.  The protocols for incubation of the bacteria can be found in
Ref.~\cite{berman21}, and the protocols for culturing of the algae can be found in Ref.~\cite{Yoest22}. 

Motion of the microbes is observed with a Nikon Eclipse inverted microscope with a 40X objective
for the bacteria and a 4X objective for the algae. For all of the microbes except the
GFP bacteria, we use diascopic illumination resulting in images where the microbes appear dark
in a bright background. For the GFP bacteria, we use epifluorescent microscopy which results
in images where the bacteria appear bright on a dark background.
Images are acquired with an Andor Zyla sCMOS
camera and digitized on a Linux workstation. The images are processed in real-time: each image
is subtracted from a background image (the subtraction is reversed for the GFP bacteria),
the subtracted image is then thresholded, and the coordinates and intensities of each pixel
whose intensity exceeds the threshold are stored in real time to the disk.  With this approach,
we can store hours of video at 30 - 50 frames per second. 

\begin{figure}
    \includegraphics[width=0.95\columnwidth]{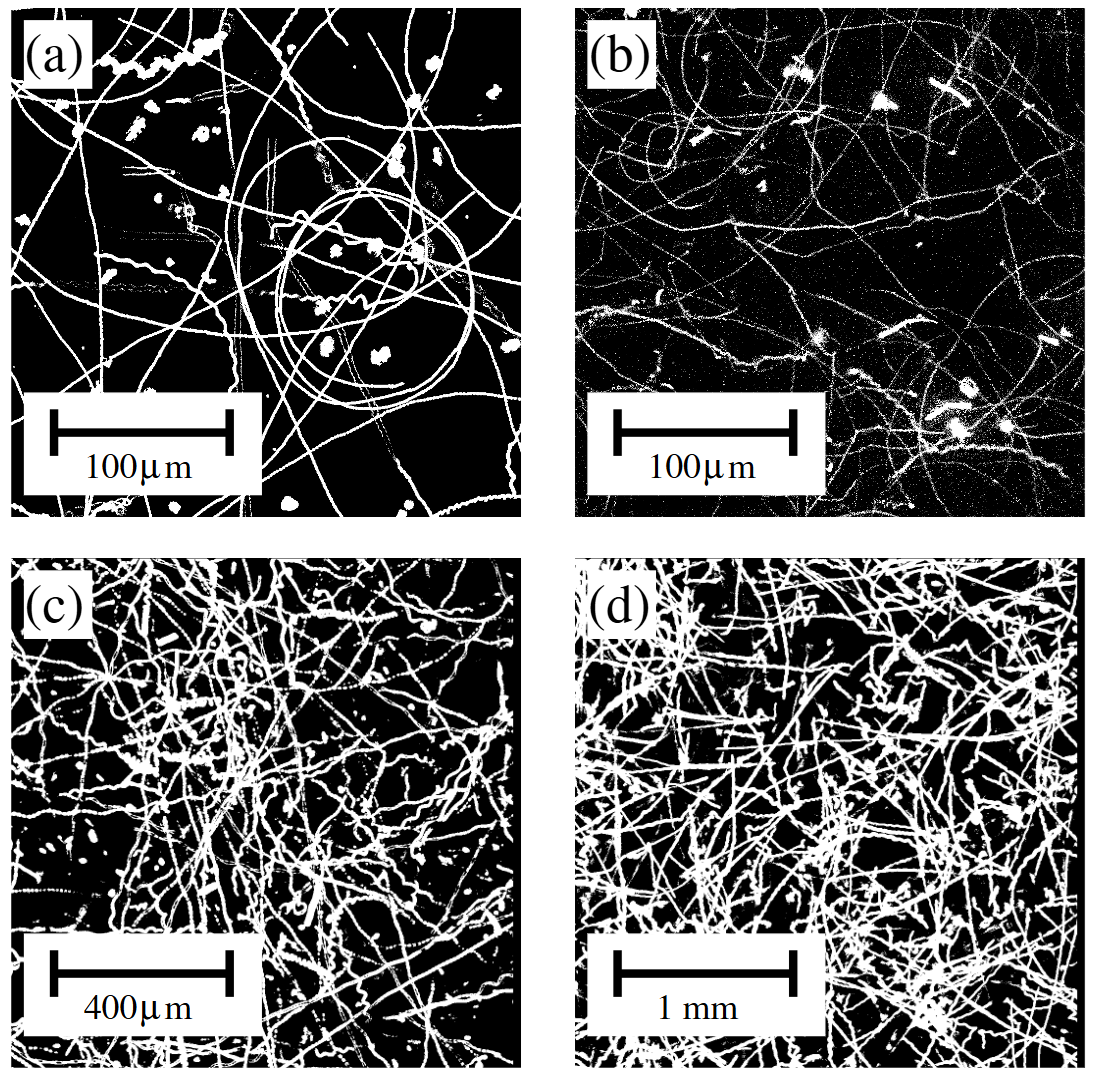}
    \caption{Streak images of the raw data for the microbes. (Videos are available in Supplementary material.)  (a) 40 s interval for smooth-swimming \textit{B. subtilis}. (b) 20 s interval for the tumbling, GFP \textit{B. subtilis}.  (c) 10 s interval for \textit{Tetraselmis}. (d) 20 s interval for \textit{Euglena}. }
    \label{fig:streaks}
\end{figure}

Examples of streak images of raw, thresholded data are shown in Fig.~\ref{fig:streaks} for each of the microbes.  (Movies of these are available in Supplementary material.)  Sessile microbes, pixel noise from the camera, and fuzz from microbes that are not in the focal plane show up in these raw images.  The range of swimming behavior is also apparent.  For example, not all of the ``smooth-swimming'' bacteria (Fig.~\ref{fig:streaks}a) swim in straight lines; some trajectories are curved, and oscillations are also visible in some trajectories, presumably a 2D projection of helical swimming. 
The raw data are processed 
in IDL with a tracking package \cite{idltracking} which determines individual
particle trajectories.  We keep only trajectories that move more than a minimum distance (typically about a fifth of the field of view), effectively removing any
sessile microbes or pixel noise from the data set. For each saved trajectory, the orientation $\theta$ for the swimming
is determined from the direction of the instantaneous velocity, determined by fitting parabolas
to the $x-$ and $y-$components of the trajectories and taking the derivative 

\begin{figure*}
    \centering
    \includegraphics[width=1.0\textwidth]{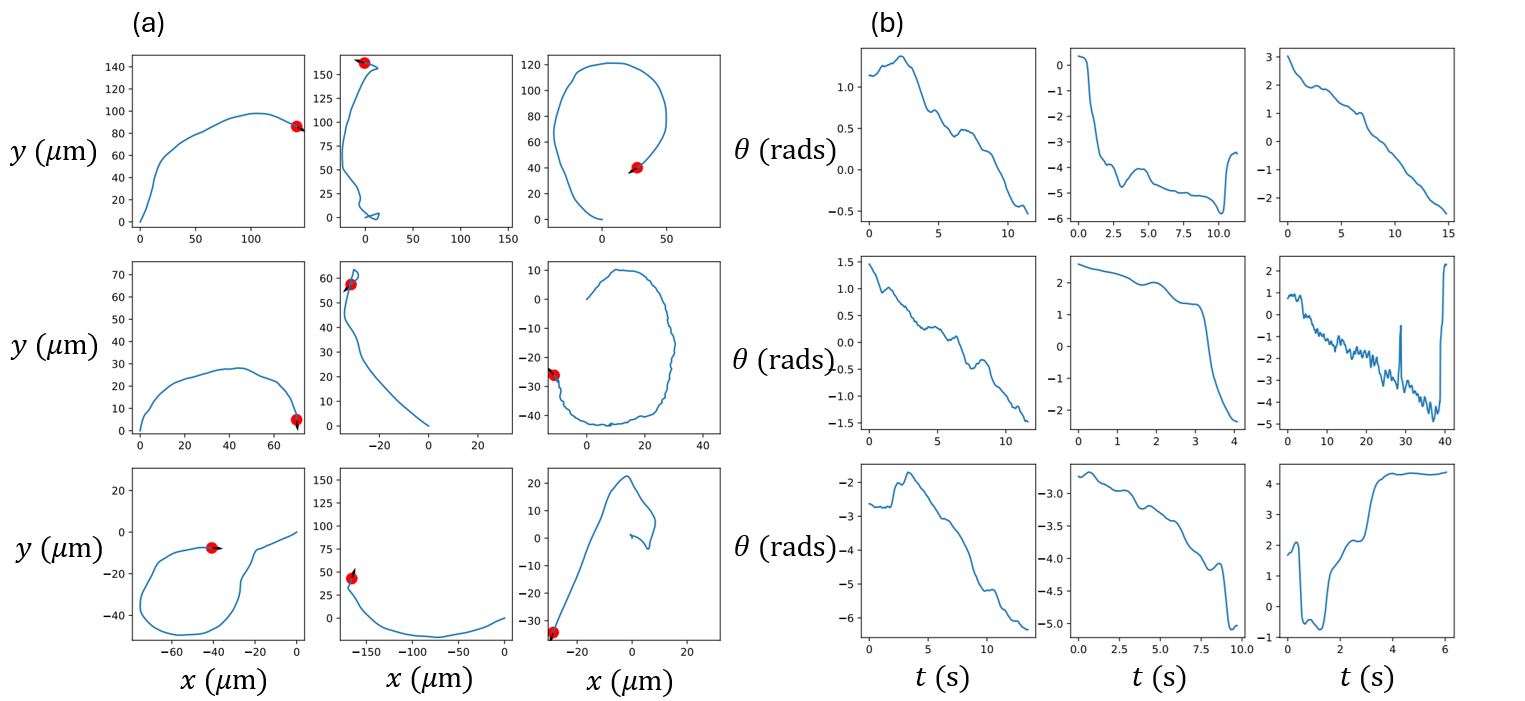}
    \caption{Example trajectories for the tumbling strain of \textit{B. subtilis}. (a) Position-space trajectories with the red dot representing the final position. There are circular arcs with different sizes, straight runs, and some tumbling events. (b) Orientation angle as a function of time corresponding to the same trajectories. The large jumps correspond to tumbling events. There appears to be constant drift (linear behavior in time) in theta for the circular trajectories.}
    \label{fig:Wild type Traj Examples}
\end{figure*}

\section{\label{sec: Stats}Orientation statistics}

\subsection{Statisical methods \label{subsec: Stats}}

\subsubsection{Expansion of delta function}

We begin our analysis of the microbial angular dynamics by constructing an initial delta function in angle and examining its subsequent expansion.  This closely resembles theoretical Green function methods for modeling diffusive systems, such as the Fokker-Planck equation. We first rectify the trajectories by mapping them from discontinuous functions on $[-\pi,\pi]$ to continuous functions on $(-\infty,\infty)$.  We next rotate all trajectories to begin at $\theta(0)=0$. We then double the ensemble using the reflection $\theta(t)\mapsto-\theta(t)$ to symmetrize the distribution. Finally, we compute histograms over $\theta$ at each time-step, normalized to unit area over the domain $(-\infty,\infty)$.

\subsubsection{Candidate distributions}

We fit the angular probability distribution and the angular velocity distribution to three candidate distributions.  The first two candidates are specific L\'{e}vy $\alpha$-stable distributions: the Lorentzian (or Cauchy) distribution (L\'{e}vy index $\alpha=1$) and the Gaussian distribution (L\'{e}vy index $\alpha=2$), given by
\begin{align}
L(x;\gamma)=\frac{1}{\pi}\frac{\gamma}{\gamma^2+x^2},\label{e1}
\end{align}
\begin{align}
G(x;\sigma)=\frac{1}{\sqrt{2\pi \sigma^2}}\exp\left(-\frac{x^2}{2\sigma^2}\right).\label{e2}
\end{align}
In Eq.~(\ref{e1}) $\gamma$ is the scale parameter, which measures the ``width" of the distribution.  The Lorentzian is characterized by heavy tails with an undefined mean and standard deviation, consistent with the presence of L\'{e}vy walks. The Lorentzian has power law tails where the tails go as $L(x)\sim x^{-2}$. In Eq.~(\ref{e2}), however, the standard deviation $\sigma$ and mean (0) are well defined, and the tails are exponential. The final candidate distribution is the Voigt profile, defined as a convolution of a Gaussian and Lorentzian 
\begin{align}
V(x;\sigma,\gamma)&=G(x;\sigma)*L(x;\gamma)\nonumber\\
&=\int_{-\infty}^{\infty}G(x';\sigma)L(x-x';\gamma)dx'\nonumber\\
&=\frac{1}{\sqrt{2\pi\sigma^2}}\text{Re}\left\{w \left( \frac{x+i \gamma}{\sqrt{2\sigma^2}}\right) \right\}, \label{e3}
\end{align}
where the Voigt profile is proportional to the real part (Re) of the Faddeeva function $w(z)=\exp(-z^2)\text{erfc}(-iz)$, with $\text{erfc}$ representing the complementary error function.  The Voigt profile is the distribution one gets by adding a Gaussian random variable with standard deviation $\sigma$ to a Lorentzian random variable with scale parameter $\gamma$. The tails at larger $x$ for the Voigt profile are dominated by the power law behavior of the Lorentzian. That is $V(x)\sim x^{-2}$ at the tails. 

\subsubsection{Fit metric}
To measure the ``goodness" of fit for a specific model of a probability distribution, we use the Akaike information criterion (AIC).  Given a set of models, this method can determine the likelihood of each model being the correct fit. There is a penalty based on the number of parameters in the model, so that simpler models with the similar accuracy are selected.  In the case of least squares regression, the AIC is defined as follows~\cite{Burnham11}
\begin{align}
\text{AIC}=n\ln\left(\frac{\text{RSS}}{n}\right)+2K,
\end{align}
 where $n$ is the sample size, RSS is the residual sum of squares for the fitted model, and $K$ is the number of parameters in the model. A more meaningful metric is to compare the AIC of all the models tested against the minimum AIC out of the full set of models, i.e. to compute
\begin{align}
\Delta_i=\text{AIC}_i-\min\{\text{AIC}_j | \forall j \}.
\end{align}
 From this value we obtain the \emph{relative} likelihood $\ell_i$ for model $i$ as
\begin{align}
\ell_i=\exp\left(-\frac{\Delta_i}{2}\right).
\end{align}

\subsection{Statistical analysis of the microbe trajectory data}

\subsubsection{\label{subsec: Wild Bac}Tumbling strain of \textit{Bacilus subtilis}}

\begin{figure}[h]
    \centering
    \includegraphics[width=1.1\columnwidth]{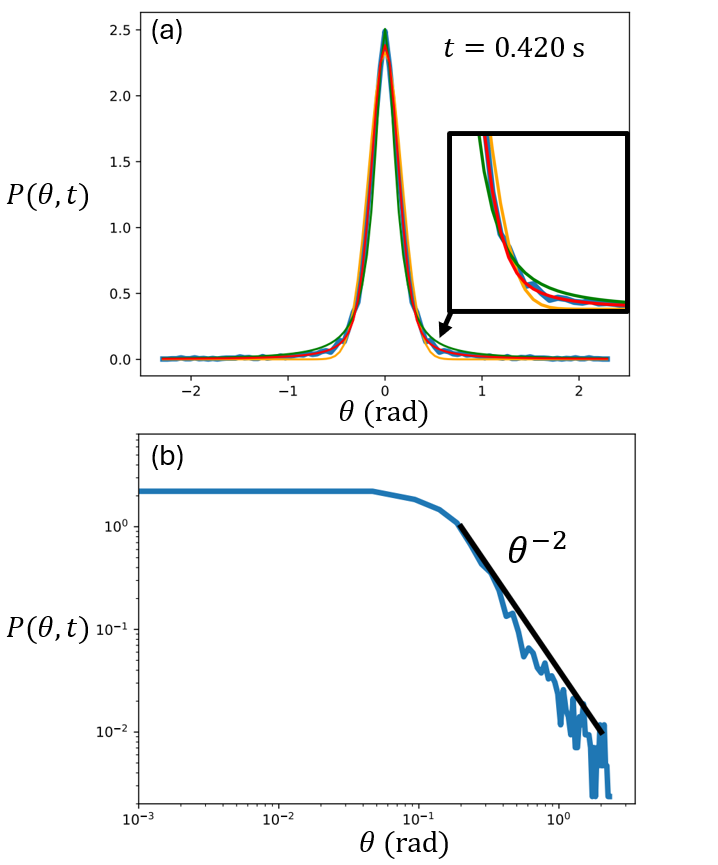}
    \caption{ (a) Expansion of the delta function centered at $\theta(0)=0$ at $t=0.420$s with the tumbling strain of \textit{B. subtilis} data in blue, the Lorentzian fit in green, the Voigt profile fit in red, and the Gaussian fit in orange. The inset is a zoomed in image of the tail of the distribution. The Voigt profile fits the data very well. (b) Log-log plot of the tails of the angular distribution with data in blue and a power law of $\theta^{-2}$ in black. The behavior of the tails is consistent with a power law for the Lorentzian or Voigt profile. }
    \label{fig:WildBac Distributions}
\end{figure}

\textit{Bacilus subtilis} (\textit{B. subtilis}) is a model rod-shaped microbe used for many scientific experiments and biophysical applications. It is one of the most studied and well-understood microbes in the scientific literature.  It is generally considered a ``good" bacterium because of its role as a probiotic gut microbe \cite{Stulke23}. It moves via the collective motion of bundles of flagellar filaments \cite{Najafi19}. An example image of the microbe is shown in Fig.~\ref{fig:Microbes}a. A streak image (Fig.~\ref{fig:streaks}b) shows the swimming behavior.  A movie of the tumbling strain of \textit{B. subtilis} swimming is provided in the supplementary material (S1). The typical spacing between \textit{B. subtilis} at any given time is at least approximately $200$ $\mu$m, or about 100 body lengths. Thus, we are in the dilute limit where the local flows should not significantly affect the statistics and we may assume the swimmers do not typically interact. 

After additional filtering to remove trajectories that might get stuck on a sessile microbe (as outlined in the supplemental material), we analyze 4547 trajectories. Example trajectories after filtering are shown in Fig.~\ref{fig:Wild type Traj Examples}. Figure~\ref{fig:Wild type Traj Examples}a shows the position-space trajectories, and Fig.~\ref{fig:Wild type Traj Examples}b shows corresponding orientation angle as a function of time. Note that many trajectories follow circular arcs with different sizes, with many showing constant drift in $\theta(t)$, except for the jumps due to large tumbling events.

\begin{table}[h]
\caption{$P(\theta,t)$ AIC for the tumbling stain of \it{B. subtilis}\label{Table:Wild Type delta func AIC}}
\begin{tabular}{ ||c | c | c | c||}

\hline
  AIC& Gaussian & Lorentzian & Voigt profile \\ 
  \hline
 $\Delta_i$& 22,900& 23,400& 0.0\\  
 \hline
 $\ell_i$& 0.0 & 0.0 & 1.0\\ 
 \hline  
\end{tabular}
\end{table}

 We begin our statistical analysis by looking at the angular probability of the ensemble as a function of time. We start with the expansion of an initial delta function centered at $\theta=0$ (as outlined in Sec.~\ref{subsec: Stats}). We then fit the candidate distributions shortly after the expansion of the delta function at $t=0.420$ s (Fig.~\ref{fig:WildBac Distributions}a).  This time was chosen so that the distribution no longer resembles a pure delta function but not so long that the data becomes excessively noisy. 
 Visually the Voigt profile seems to be the better fit. Measuring the goodness of the fit more quantitatively with the AIC  in Table~\ref{Table:Wild Type delta func AIC}, the Voigt profile is clearly the best fit. The Voigt profile parameters for this fit are summarized in Table~\ref{Table:Wild Type Stats}. The Gaussian scale parameter $\sigma$ is roughly twice the Lorentzian scale parameter $\gamma$. Figure~\ref{fig:WildBac Distributions}b shows the power law behavior of the tails which is very close to $\theta^{-2}$, this is consistent with either a Lorentzian distribution or a Voigt profile. 

\begin{figure}
    \centering
    \includegraphics[width=1.08\columnwidth]{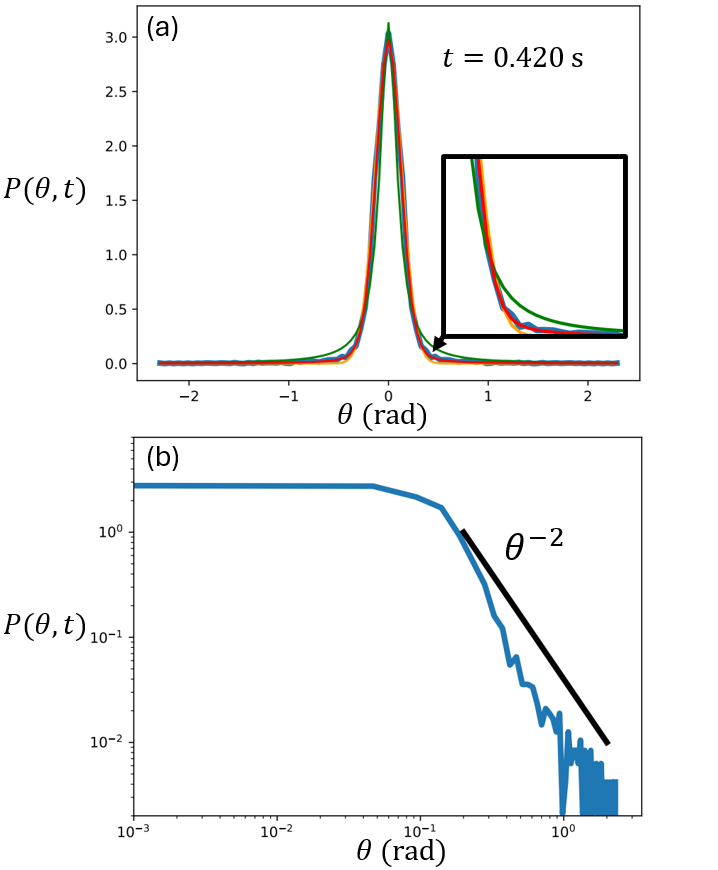}
    \caption{ (a) Expansion of the delta function centered at $\theta(0)=0$ at $t=0.420$s with the smooth swimming strain of \textit{B. subtilis} data in blue, the Lorentzian fit in green, the Voigt profile fit in red, and the Gaussian fit in orange. The inset is a zoomed in image of the tail of the distribution. The Voigt profile fits the data very well.  (b) Log-log plot of the tails of the angular distribution with data in blue and a power law of $\theta^{-2}$ in black. The behavior of the tails is close to the power law for the Lorentzian or Voigt profile. }
    \label{fig:SmoothBac Distributions}
\end{figure}

\begin{figure*}
    \centering
    \includegraphics[width=1.0\textwidth]{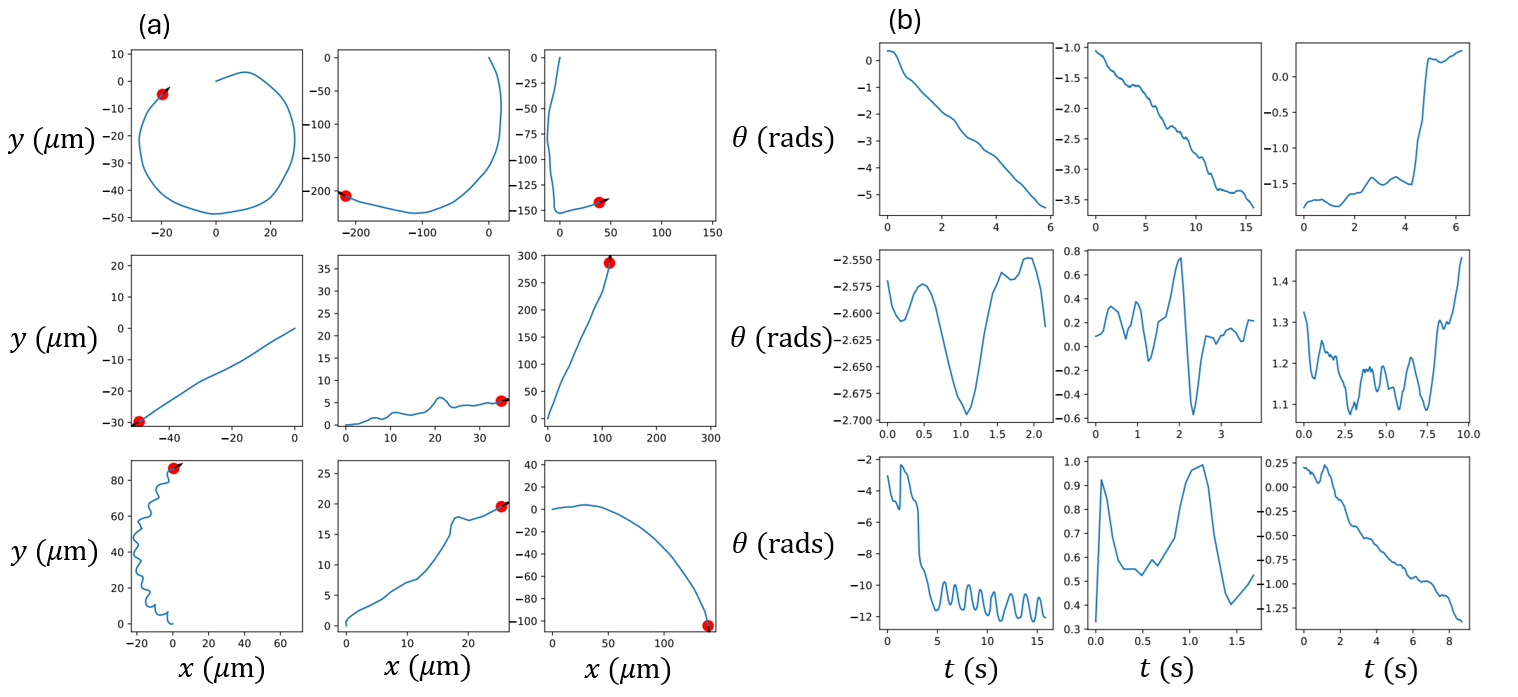}
    \caption{Example trajectories for the ``smooth" swimming strain of \textit{B. subtilis}. (a) Position space trajectories with the red dot representing the final position. There are circular arcs of different sizes, straight runs, and some tumbling events. (b) Orientation angle as a function of time corresponding to the same trajectories. The large jumps correspond to tumbling events. There appears to be constant drift (linear behavior in time) in theta for the circular trajectories.}
    \label{fig:Smooth Bac Traj Examples}
\end{figure*}

\begin{table}
\caption{Voigt profile parameters for the tumbling strain of \it{B. subtilis} \label{Table:Wild Type Stats}}
\begin{tabular}{ ||c | c ||}

\hline
  Voigt Parameters& $P(\theta,t)$  \\ 
  \hline
 $\sigma$ & $0.109 \pm 0.002$ rad\\  
 \hline
 $\gamma$ & $0.067 \pm 0.002$ rad\\ 
 \hline  
\end{tabular}
\end{table}

\subsubsection{\label{subsec: Smooth Bac}Smooth-swimming strain of \textit{Bacilus subtilis}}

In addition to the tumbling strain of \textit{B. subtilis}, we also analyze the statistics for a ``smooth" swimming strain of \textit{B. subtilis} outlined in Sec.~\ref{sec: Experiment}. A video of smooth swimming \textit{B. subtilis} is contained in the supplementary material (S2). A streak image (Fig.~\ref{fig:streaks}a) shows the swimming behavior. The volume fraction for the smooth swimming strain of \textit{B. subtilis} is roughly similar to the tumbling strain. Thus, we assume the swimmers do not interact with each other.

After filtering out the sessile microbes from the statistics as outlined in the supplementary material, we analyze 5100 trajectories. Example trajectories after filtering are shown in Fig.~\ref{fig:Smooth Bac Traj Examples}. Figure~\ref{fig:Smooth Bac Traj Examples}a shows the position space trajectories, and Fig.~\ref{fig:Smooth Bac Traj Examples}b shows the corresponding orientation angle as a function of time. The trajectories are very similar to the tumbling strain, namely the presence of circular arcs. There are also helices, straight runs, and tumbling events.  

\begin{table}
\caption{$P(\theta,t)$ AIC for the smooth swimming strain of \it{B. subtilis}\label{Table:Smooth Bac delta func AIC}}
\begin{tabular}{ ||c | c | c | c||}

\hline
  AIC& Gaussian & Lorentzian & Voigt profile \\ 
  \hline
 $\Delta_i$& 12,400& 32,400& 0.0\\  
 \hline
 $\ell_i$& 0.0 & 0.0 & 1.0\\ 
 \hline  
\end{tabular}
\end{table}

We begin our analysis with the expansion of the delta function for the angular probability as a function of time. We fit the candidate distributions to the expansion of the delta function at the same time-step as the tumbling strain at $t=0.420$ s (Fig.~\ref{fig:SmoothBac Distributions}a). Visually the Voigt profile is a better fit. Using the AIC summarized in Table~\ref{Table:Smooth Bac delta func AIC}, we conclude that the Voigt profile is clearly the better fit. The Voigt profile parameters for this fit are given in Table~\ref{Table:Smooth Bac Stats}. The Lorentzian component of the Voigt profile is much smaller than the Gaussian component. This strain of \textit{B. subtilis} tends to swim more smoothly than the tumbling strain. If we compare the Lorentzian width to the tumbling strain it is roughly half the size. If the Lorentzian distribution describes tumbling dynamics, then we expect smoother swimmers to have a smaller Lorentzian part, as seen here. Figure~\ref{fig:SmoothBac Distributions}b shows the power law behavior of the tails which is very close to $\theta^{-2}$, consistent with a Lorentzian or Voigt profile. 

\begin{table}[h]
\caption{Voigt profile parameters for smooth swimming strain of \it{B. subtilis} \label{Table:Smooth Bac Stats}}
\begin{tabular}{ ||c | c ||}

\hline
  Voigt Parameters& $P(\theta,t)$\\ 
  \hline
 $\sigma$ & $0.114 \pm 0.002$ rad\\  
 \hline
 $\gamma$ & $0.025 \pm 0.002$ rad\\ 
 \hline  
\end{tabular}
\end{table}

\begin{figure}
    \centering
    \includegraphics[width=1.05\columnwidth]{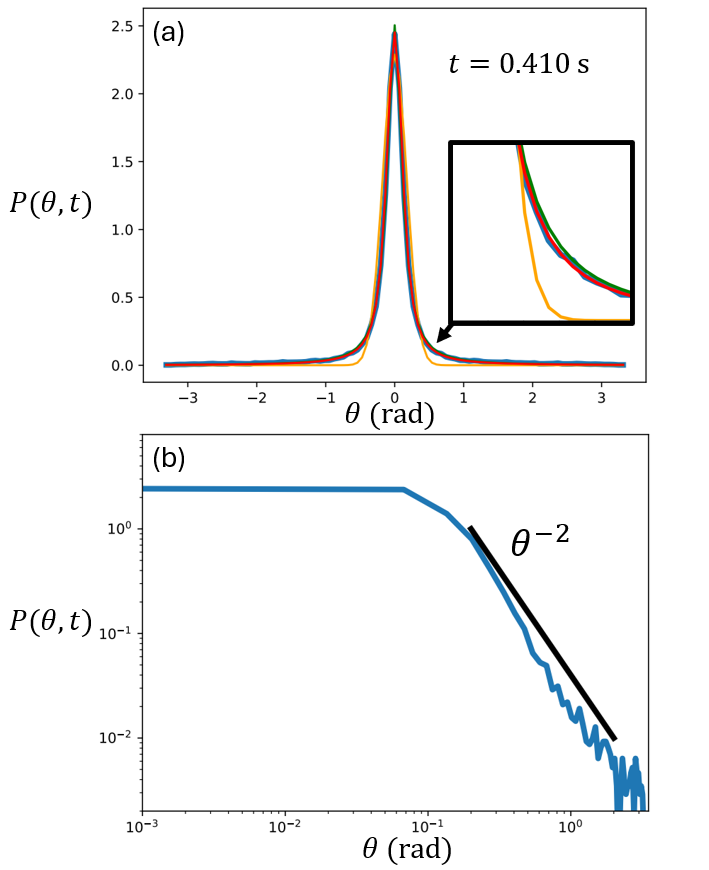}
    \caption{ (a) Expansion of the delta function centered at $\theta(0)=0$ at $t=0.420$s with \textit{Tetraselmis} data in blue, the Lorentzian fit in green, the Voigt profile fit in red, and the Gaussian fit in orange. The inset is a zoomed in image of the tail of the distribution. The Voigt profile and Lorentzian are very close to each other. However, the Voigt profile fits the tails slightly better.  (b) Log-log plot of the tails of the angular distribution with data in blue and a power law of $\theta^{-2}$ in black. The behavior of the tails is consistent with a power law for the Lorentzian or Voigt profile. }
    \label{fig:Tetra Distributions}
\end{figure}

\begin{figure*}
    \centering
    \includegraphics[width=1.0\textwidth]{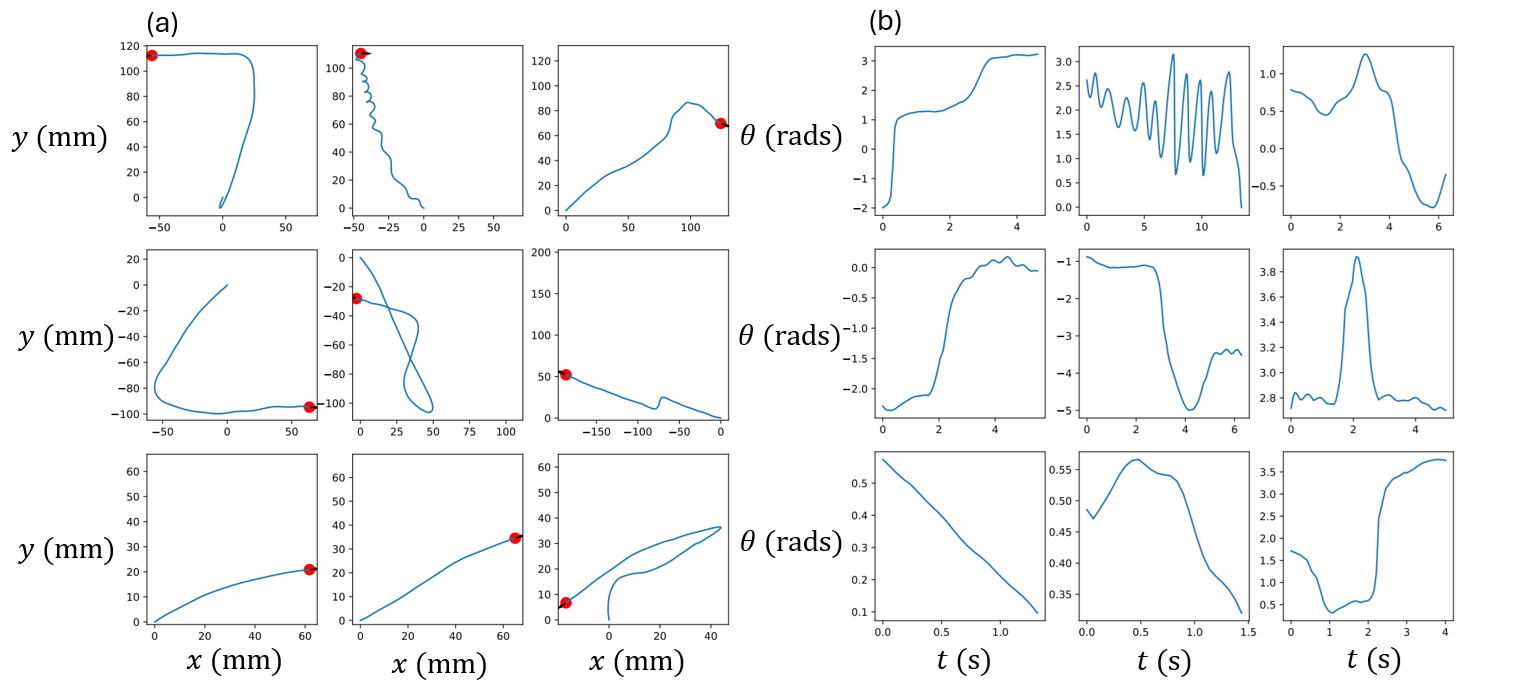}
    \caption{Example trajectories for \textit{Tetraselmis}. (a) Position space trajectories with the red dot (with orientation in black) representing the final position. There are many straight runs, tumbling events, and some helices. (b) Orientation angle as a function of time corresponding to the same trajectories. The large jumps correspond to tumbling events. }
    \label{fig:Tetra Traj Examples}
\end{figure*}

\subsubsection{\label{subsec: Tetraselmis}\textit{Tetraselmis suecica}}

\textit{Tetraselmis suecica} is a motile Eukaryotic marine algae (Fig.~\ref{fig:Microbes}b).  It moves via a breast stroke-like motion with four flagella, which pull fluid in front of the cell body~\cite{Wan20}. \textit{Tetraselmis} has been studied as a model marine algae for sustainability and environmental applications, such as waste water remediation and bulk chemical production \cite{Mehariya24}.  A video of swimming \textit{Tetraselmis} is included in the supplementary material (S3). A streak image (Fig.~\ref{fig:streaks}c) shows the swimming behavior. The typical spacing between \textit{Tetraselmis} in the field of view is at least approximately 200 $\mu$m or about 15–20 body lengths. Thus, the local flows are negligible, and the swimmers do not typically interact. 

After filtering out sessile microbes as outlined in the supplementary material, we analyze the angular statistics for 12,758 trajectories. Example trajectories after filtering are shown in Fig.~\ref{fig:Tetra Traj Examples}. Figure~\ref{fig:Tetra Traj Examples}a shows the position space trajectories, and Fig.~\ref{fig:Tetra Traj Examples}b shows the corresponding orientation angle as a function of time. The trajectories follow helices, straight runs, and have clear dramatic tumbling events.

We begin our analysis of the angular statistics via the expansion of the angular probability from a delta function. We fit the candidate distributions at the time-step $t=0.410$ s (a similar time to the previous microbial populations, see Fig.~\ref{fig:Tetra Distributions}a). Visually the Voigt profile and Lorentzian distributions are indistinguishable in the main image. However looking at the inset of Fig.~\ref{fig:Tetra Distributions}a, the Voigt profile fits the tails slightly better. Using the AIC to measure the ``goodness" of fit (summarized in Table.~\ref{Table:Tetra delta func AIC}), the Voigt profile is clearly the better fit, although it appears the Lorentzian is a better fit that the Gaussian. The Voigt profile parameters for this fit are given in Table~\ref{Table:Tetra Stats}. The Lorentzian part of the Voigt profile is approximately twice the width of the Gaussian part. The Lorentzian part of the Voigt profile dominates the distribution. This is expected if we assume the Lorentzian part of the distribution describes tumbling. \textit{Tetraselmis} is larger than \textit{B. subtilis}. Thus, the rotational diffusion constant due to the temperature and viscosity of the fluid is smaller which may contribute to the Gaussian part of the Voigt profile being less significant.  Figure~\ref{fig:Tetra Distributions}b shows the power law behavior of the tails which is very close to $\theta^{-2}$, consistent with a Lorentzian or Voigt profile. 

\begin{table}[h]
\caption{$P(\theta,t)$ AIC for \it{Tetraselmis}\label{Table:Tetra delta func AIC}}
\begin{tabular}{ ||c | c | c | c||} 

\hline
  AIC& Gaussian & Lorentzian & Voigt profile \\ 
  \hline
 $\Delta_i$& 336,000& 143,000& 0.0\\  
 \hline
 $\ell_i$& 0.0 & 0.0 & 1.0\\ 
 \hline  
\end{tabular}
\end{table}

\begin{table}[h]
\caption{Voigt profile parameters \it{Tetraselmis} \label{Table:Tetra Stats}}
\begin{tabular}{ ||c | c ||}

\hline
  Voigt Parameters& $P(\theta,t)$ \\ 
  \hline
 $\sigma$ & $0.0526 \pm 0.0006$ rad\\  
 \hline
 $\gamma$ & $0.1107 \pm 0.0003$ rad\\ 
 \hline  
\end{tabular}
\end{table}

\begin{figure*}
    \centering
    \includegraphics[width=1.0\textwidth]{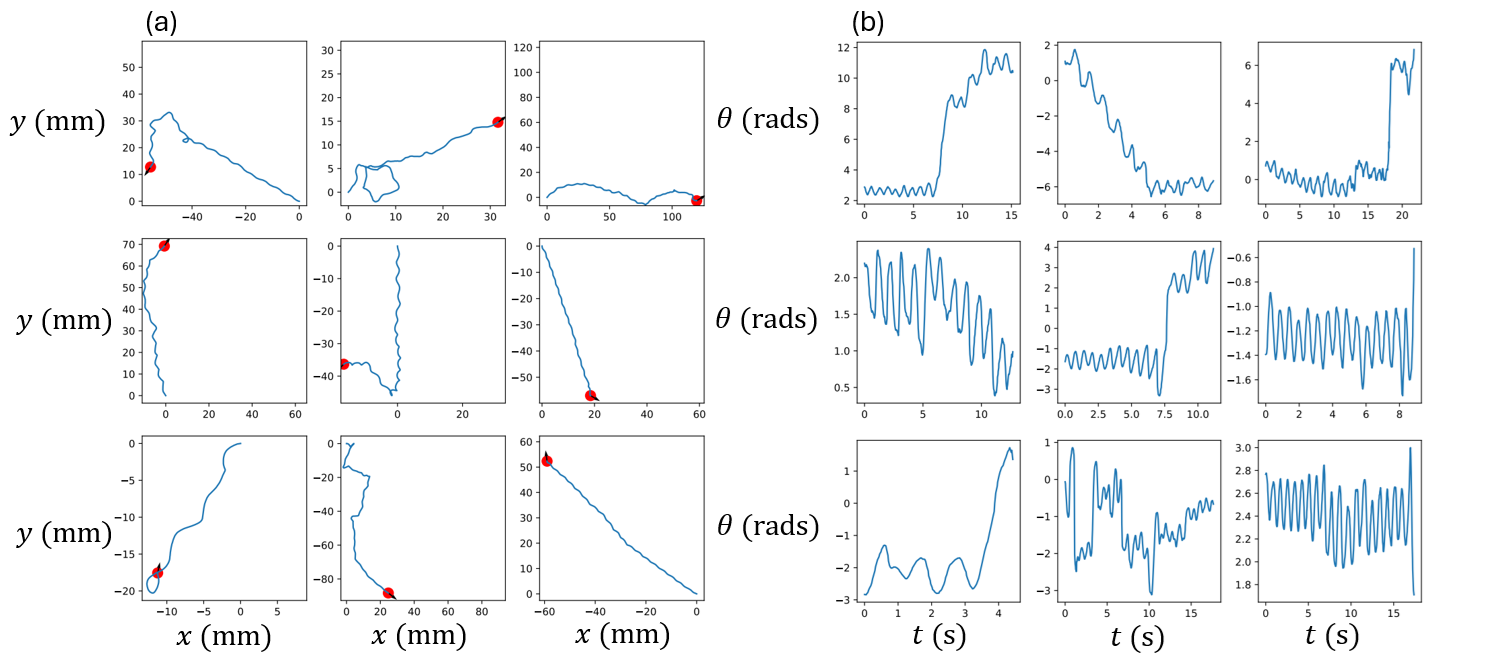}
    \caption{ Example trajectories for \textit{Euglena}. (a) Position space trajectories with the red dot (with orientation in black) representing the final position. Virtually all trajectories in this ensemble are helices, there are also clear tumbling events. (b) Orientation angle as a function of time corresponding to the same trajectories. The large jumps correspond to tumbling events. Notice how most of the runs are sinusoidal in $\theta(t)$.}
    \label{fig:Euglena Traj Examples}
\end{figure*}

\subsubsection{\label{subsec: Euglena}\textit{Euglena gracilis}}

\textit{Euglena gracilis} is a mixotrophic algae. Its primary food source is light. It possesses two flagella (dorsal and ventral). Only the dorsal flagellum leaves the cell. \textit{Euglena}'s cell surface architecture allows it to exhibit a wide range of swimming behavior \cite{Konur20}. Its average mode of locomotion is an off-axis puller~\cite{Giuliani21}, where it uses its whip-like flagella to pull fluid in front of the cell body. Current research of \textit{Euglena} points towards possible applications as a source of dietary supplements~\cite{Gissibl19}. A picture of \textit{Euglena} can be seen in Fig.~\ref{fig:Microbes}c. A video of a collection of \textit{Euglena} swimming can be seen in the supplementary material (S4). A streak image (Fig.~\ref{fig:streaks}d) shows the swimming behavior.  The spacing between \textit{Euglena} is approximately 250 $\mu$m (in the worst case scenario), which is 3-5 body lengths. Thus, \textit{Euglena} may weakly interact in some cases.

This population of microbes had virtually no sessile trajectories. Thus no filtering was needed. We analyze the angular statistics for 13,454 trajectories. Figure~\ref{fig:Euglena Traj Examples} shows example trajectories for this population of \textit{Euglena}. Figure~\ref{fig:Euglena Traj Examples}a shows position-space trajectories. Virtually all trajectories in this ensemble are helical during runs, with many clear tumbling events. Figure~\ref{fig:Euglena Traj Examples}b shows the corresponding orientation angle as a function of time. The runs are sinusoidal as expected from helices. The large jumps in $\theta$ correspond to tumbling events. 

\begin{table}[h]
\caption{$P(\theta,t)$ AIC for \it{Euglena}\label{Table:Euglena delta func AIC}}
\begin{tabular}{ ||c | c | c | c||}

\hline
  AIC& Gaussian & Lorentzian & Voigt profile \\ 
  \hline
 $\Delta_i$& 27,600& 24,900& 0.0\\  
 \hline
 $\ell_i$& 0.0 & 0.0 & 1.0\\ 
 \hline  
\end{tabular}
\end{table}

\begin{figure}
    \centering
    \includegraphics[width=1.05\columnwidth]{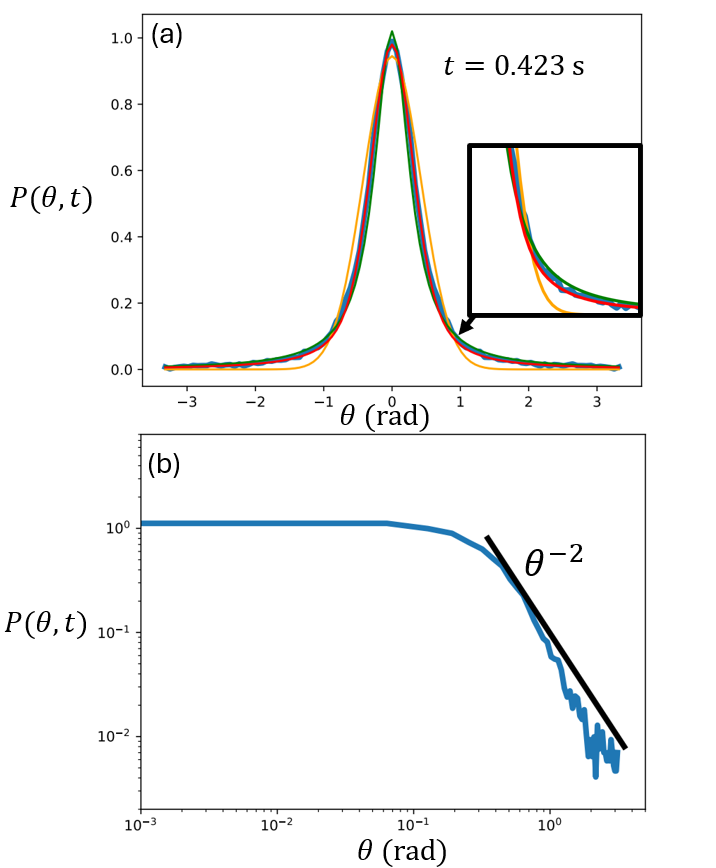}
    \caption{ (a) Expansion of the delta function centered at $\theta(0)=0$ at $t=0.423$s with \textit{Euglena} data in blue, the Lorentzian fit in green, the Voigt profile fit in red, and the Gaussian fit in orange. The inset is a zoomed in image of the tail of the distribution. The Voigt profile and Lorentzian are very close to each other. However, the Voigt profile fits the peak and tails slightly better.  (b) Log-log plot of the tails of the angular distribution with data in blue and a power law of $\theta^{-2}$ in black. The behavior of the tails is consistent with a power law for the Lorentzian or Voigt profile. }
    \label{fig:Euglena Distributions}
\end{figure}

We begin our analysis of the angular statistics via the expansion of the delta function for the angular probability. We fit the candidate distributions at $t=0.423$s (a similar time to the other microbes, see Fig.~\ref{fig:Euglena Distributions}a). The Lorentzian and Voigt profile fits are very close to each other. However the Voigt profile fits the tails and peak better. Using the AIC (Table~\ref{Table:Euglena delta func AIC}), we conclude the Voigt profile is clearly the best fit. Table~\ref{Table:Euglena Stats} shows the Voigt profile parameters for this fit. Both the Gaussian and Lorentzian parts of the Voigt profile are similar in scale. \textit{Euglena} is very large compared to the other microbes. Thus, we expect the rotational diffusion constant to be negligible. Therefore, the relatively large Gaussian part of the Voigt profile may be caused by some process other than physical rotational diffusion, which we will discuss in the next section. Figure~\ref{fig:Euglena Distributions}b shows the power law behavior of the tails which is very close to $\theta^{-2}$, consistent with a Lorentzian or Voigt profile. 

\begin{table}
\caption{Voigt profile parameters for \it{Euglena} \label{Table:Euglena Stats}}
\begin{tabular}{ ||c | c ||}

\hline
  Voigt Parameters& $P(\theta,t)$ \\ 
  \hline
 $\sigma$ & $0.219 \pm 0.003$ rad\\  
 \hline
 $\gamma$ & $0.207 \pm 0.002$ rad\\ 
 \hline  
\end{tabular}
\end{table}

\section{\label{sec: Theory}Theoretical models for the time-evolution of angular probability distributions}

\subsection{\label{sec: SDE/FPKE}Stochastic ODE and Fokker-Planck equation}

We first consider a general one dimensional stochastic differential equation (SDE) for the orientational dynamics in two dimensions
\begin{align}
\dot{\theta}(t)=f(\theta,t)+\xi(t),
\end{align}
where $f(\theta,t)$ is known as the drift term, and $\xi(t)$ is the noise term. If this system begins at initial value $\theta(0)=0$, then the probability distribution $P(\theta,t)$ is given by a Fokker-Planck equation
\begin{align}
\frac{\partial}{\partial t}P(\theta,t)=-\frac{\partial}{\partial \theta} \left[f(\theta,t)P(\theta,t) \right]+\mathcal{L}P(\theta,t), \label{FPKE}
\end{align}
with initial condition $P(\theta,t=0)=\delta(\theta)$. The operator $\mathcal{L}$ is determined by the noise term $\xi(t)$. Now let us consider the case with an additive combination of Gaussian ($\xi_G(t)$) and Lorentzian ($\xi_L(t)$) noise.  Then the SDE becomes \cite{Zan20}
\begin{align}
\dot{\theta}(t)=f(\theta,t)+\xi_G(t)+\xi_L(t),
\end{align}
and the probability distribution evolves as \cite{Zan20}
\begin{align}
\frac{\partial}{\partial t}P(\theta,t)&=-\frac{\partial}{\partial \theta}[f(\theta,t)P(\theta,t)]\nonumber\\ 
&+D_G\frac{\partial^2}{\partial \theta^2}P(\theta,t)+D_L\frac{\partial}{\partial|\theta|}P(\theta,t),\label{L+G}
\end{align}
where $D_G$ and $D_L$ denote the Gaussian and Lorentzian noise strength respectively. The derivative $\partial / \partial|\theta|$ denotes the Riesz fractional derivative which can be thought of as a fractional Laplacian. The order of the Riesz fractional derivative is the L\'{e}vy index for L\'{e}vy-$\alpha$ stable noise terms. For a L\'{e}vy index of $\alpha=1$ it is defined in one of two ways, the first being 
\begin{align}
\frac{d}{d|x|}g(x)&=-\frac{d}{dx}[Hg(x)],\\
Hg(x)&=\frac{1}{\pi}\int_{-\infty}^{\infty}\frac{g(x')}{x-x'}dx',
\end{align}
where $H$ is the Hilbert transform operator (with integral interpreted in the principle value sense) \cite{Zan20}.   The second definition is via a Fourier transform
\begin{align}
\mathcal{F}\left\{\frac{d}{d|x|}g(x)\right\}(k)=-|k|\Tilde{g}(k),
\end{align}
where $\Tilde{g}(k)$ is the Fourier transform of $g(x)$.

The advantage of Eq.~(\ref{L+G}) is that it is analytically solvable in the case of no drift (i.e. $f(\theta,t)=0$). The solution can be found via Fourier transforms to obtain a Voigt profile~\cite{Zan20}
\begin{align}
&P(\theta,t)=V\left(\theta;\sigma=\sqrt{2 D_G t},\gamma=D_L t\right), 
\end{align}
where the function $V$ is given by Eq.~\ref{e3}. Note if $D_G=0$ we obtain a Lorentzian whose $\gamma$ parameter grows linearly with time. On the other hand, if $D_L=0$ we obtain the well known solution to the diffusion equation, a Gaussian whose standard deviation grows proportional to $\sqrt{t}$~\cite{Zan20}.

There is a timescale associated with the solution to the Fokker-Planck equation where the Voigt profile switches from Gaussian dominated to Lorentzian dominated. It is defined as the time where $\sigma=\gamma$
\begin{align}
    t_L=\frac{2D_G}{D_L^2}.
\end{align}
At times $t<t_L$ the distribution is always Gaussian dominated. This is contrasted by times $t>t_L$ where the distribution is always Lorentzian dominated. We can think of $t_L$ as defining the time it takes for the distribution to become more heavy tailed and have a relatively higher probability of  making larger jumps in angle.

\subsection{\label{sec: Ensemble Theory}Ensemble theory for random drift parameters}

Inspired by the behavior of some of the trajectories, such as helices with different sizes and frequencies, or circular arcs with different radii, we provide a way to compute the probability distribution for ensembles of non-interacting swimmers. The stochastic system at a given vector of drift parameter values $\mathbf{a}=(a_1,...,a_k)$ is given by
\begin{align}
\dot{\theta}(t;\mathbf{a})=f(\theta,t;\mathbf{a})+\xi(t),
\end{align}
where $\xi(t)$ denotes the total noise contribution. If this system consists of an infinite amount of non-interacting particles, its probability distribution evolves according to 
\begin{align}
\frac{\partial}{\partial t}P(\theta,t;\mathbf{a})=-\frac{\partial}{\partial \theta}[f(\theta,t;\mathbf{a})P(\theta,t;\mathbf{a})]+\mathcal{L}P(\theta,t;\mathbf{a}),
\end{align}
where $\mathcal{L}$ is an operator that is determined by the nature of the noise term $\xi(t)$. Suppose now that we have an ensemble of non-interacting agents with fixed parameter values (for each trajectory) randomly chosen from the probability distribution $P_\mathbf{a}(\mathbf{a})$. Then we may find the total probability distribution $P(\theta,t)$ of the full ensemble by finding a solution to the Fokker-Planck equation $P(\theta,t;\mathbf{a})$ and summing those solutions weighted by the probability of having the vector of parameters $\mathbf{a}$, i.e. we compute the integral over all of parameter space
\begin{align}
P(\theta,t)=\int_{V_\mathbf{a}}P_\mathbf{a}(\mathbf{a})P(\theta,t;\mathbf{a})d^k\mathbf{a}.
\end{align}
where $d^k\mathbf{a}$ is the volume element in the $k$-dimensional parameter space $V_\mathbf{a}$. Example solutions to several ensembles are shown in the supplementary material. For systems with multiple random drift parameters, the integral becomes very expensive to solve both analytically and numerically.

\subsection{Applications of the theory\label{Sec:Applications}}

In this section we outline applications for the ensemble theory from Sec.~\ref{sec: Ensemble Theory} that have clear physical relevance. Additional applications and more detailed methodologies are outlined in the supplementary material.  

\subsubsection{Ensembles with distributions of circular arcs\label{Sec:Circular Arcs}}

The most noteworthy aspect of both the tumbling and smooth swimming strains of \textit{B. subtilis} is that they often move in circular arcs with different sizes. If the swimmer undergoes a Voigt random walk in addition to circular arcs we can represent angular dynamics with the equation
\begin{align}
\dot{\theta}(t)=\omega+\xi_V(t)\label{ballistic angle},
\end{align}
where $\xi_V(t)$ is an additive combination of Gaussian and Lorentzian noise, and $\omega$ is a constant drawn randomly from the probability distribution $P_\omega(\omega)$. Let us assume that $\omega$ is Gaussian distributed and centered at 0. Then 
\begin{align}
P_\omega(\omega)=G(\omega;\sigma=\sigma_\omega).
\end{align}
The solution to the Fokker-Planck equation corresponding to Eq.~(\ref{ballistic angle}) is a moving Voigt profile that expands with time
\begin{align}
    P(\theta,t;\omega)=V(\theta-\omega t;\sigma=\sqrt{2 D_G t},\gamma=D_L t)
\end{align}
We compute the integral over the whole ensemble as described in Sec.~\ref{sec: Ensemble Theory} (see supplemental material for more detail).
Using a simple substitution $y=\omega t$ we may rewrite the integral as a convolution
\begin{align}
P(\theta,t)=G(\theta;\sigma=\sigma_\omega t)*V(\theta;\sigma=\sqrt{2 D_G t},\gamma=D_L t).
\end{align}
Using the properties of convolutions we obtain
\begin{align}
P(\theta,t)=V\left(\theta;\sigma=\sqrt{\sigma_\omega^2t^2+2 D_G t},\gamma=D_L t\right),\label{ballistic eq}
\end{align}
 which means the random distribution of ballistic angles makes $\sigma^2$ a polynomial in time with both a quadratic and a linear term.

\begin{figure*}
    \centering
    \includegraphics[width=1.0\textwidth]{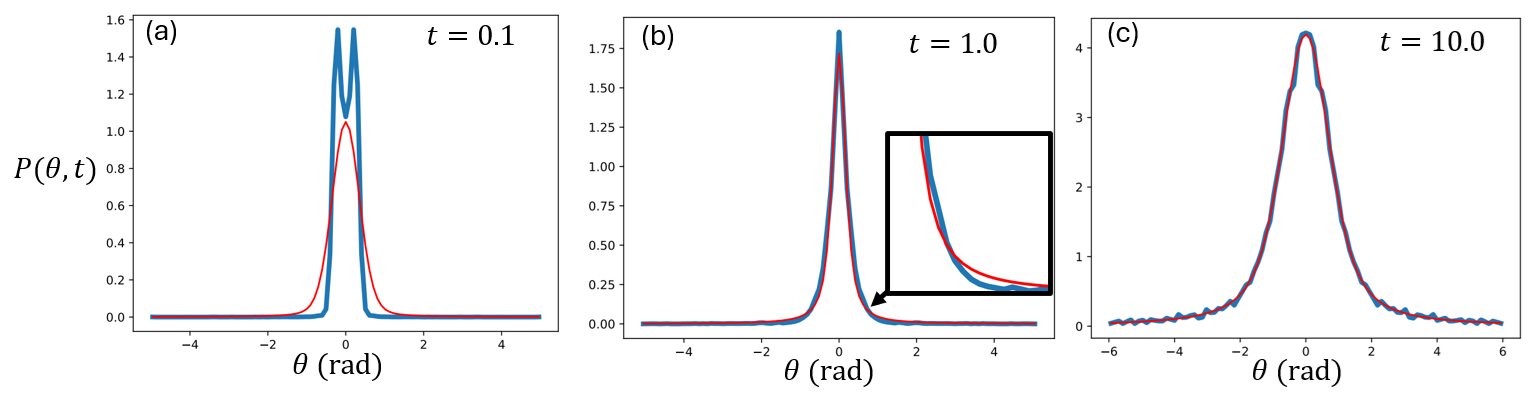}
    \caption{Distributions for the ensemble of helical trajectories with $D_G=0$, $D_L=0.05$, $\mu_A=0.5$, $\sigma_A=0.1$, $\mu_\omega=2\pi$, and $\sigma_\omega=0.75$.  The Monte Carlo simulation is in blue and the Voigt profile fit is in red.  (a) $t=0.1$. At early times, the distribution is not a Voigt profile, but splits apart. This is similar to an ensemble that only has variations in frequency as outlined in the supplementary material (S5 for movie). (b) $t=1.0$.  At intermediate times, the distribution becomes more of a Voigt profile. However, the inset shows that it does not quite match the tails. (c) $t=10.0$.  At late times, the Voigt profile is an excellent fit.}
    \label{fig:HelicalTheory Example Distributions}
\end{figure*}
 
 The quadratic coefficient for $\sigma^2$ is $\sigma_\omega^2$. Physically it represents the variance in the ballistic angular velocity across the ensemble, which leads to the distribution of circular arcs. The linear term in $\sigma^2$ has coefficient $2D_G$, related to the Gaussian rotational noise, most likely caused by physical rotational diffusion due to the temperature and viscosity of the fluid. The parameter $\gamma$ in this ensemble remains unchanged. It is linear with slope $D_L$ (the Lorentzian noise strength) and represents the L\'{e}vy walks in angle.

The timescales associated with this system are different than the time scale $t_L$ of the Fokker-Planck equation without the distribution of ballistic angles. The timescale where the distribution switches from Gaussian dominated to Lorentzian dominated is
\begin{align}
    t_L=\frac{2D_G}{D_L^2-\sigma_\omega^2}.
\end{align}
The distribution only switches to Lorentzian dominated if $D_L>\sigma_\omega$, otherwise it is always Gaussian dominated. Meaning if the standard deviation in the ballistic angle is large relative to the Lorentzian noise strength, then the distribution will always be Gaussian dominated.

Refering back to Eq.~(\ref{ballistic eq}), there is another timescale, where the Gaussian part of the Voigt profile switches from noise dominated to dominated by the distribution of ballistic angles
\begin{align}
    t_B=\frac{2D_G}{\sigma_\omega^2}.
\end{align}
If $t<t_B$ the Gaussian part of the Voigt profile is dominated by noise, otherwise it is dominated by the ballistic motion in $\theta$.

\subsubsection{Ensembles with distributions of helices}

A more complicated case that we expect in the data for \textit{Euglena} is a distribution of frequencies, amplitudes, and phases, i.e. to solve the SDE
\begin{align}
    \dot{\theta}(t)=A\omega\cos(\omega t -\delta)+\xi_V(t).
\end{align}

A natural choice is to draw the frequencies and amplitudes from different Gaussian distributions (which are not centered at 0), and then draw the phase from a uniform distribution over one period (since the microbes will enter the focal plane of the microscope at different times). That is
\begin{align}
P_\omega(\omega)=G(\omega-\mu_\omega;\sigma=\sigma_\omega),\\
P_A(A)=G(A-\mu_A;\sigma=\sigma_A),\\
P_\delta(\delta)=\frac{1}{2\pi}.
\end{align}
It is impractical to attempt to compute the total probability for this ensemble analytically.  Thus, we seek a numerical solution. A simple starting point is to generate Monte Carlo simulations as outlined in the supplemental material, or we can solve the three-dimensional integral numerically (which is very computationally expensive). We can fit a Voigt profile to the corresponding distribution as an approximation and analyze the behavior of the fit parameters as a function of time. To model \textit{Euglena}, we expect $D_G\approx0$, due to the size of the organism. Therefore, the angular distribution without Gaussian noise is a good way to predict what we expect to see in the \textit{Euglena} data. 

\begin{figure}
    \centering
    \includegraphics[width=1.0\columnwidth]{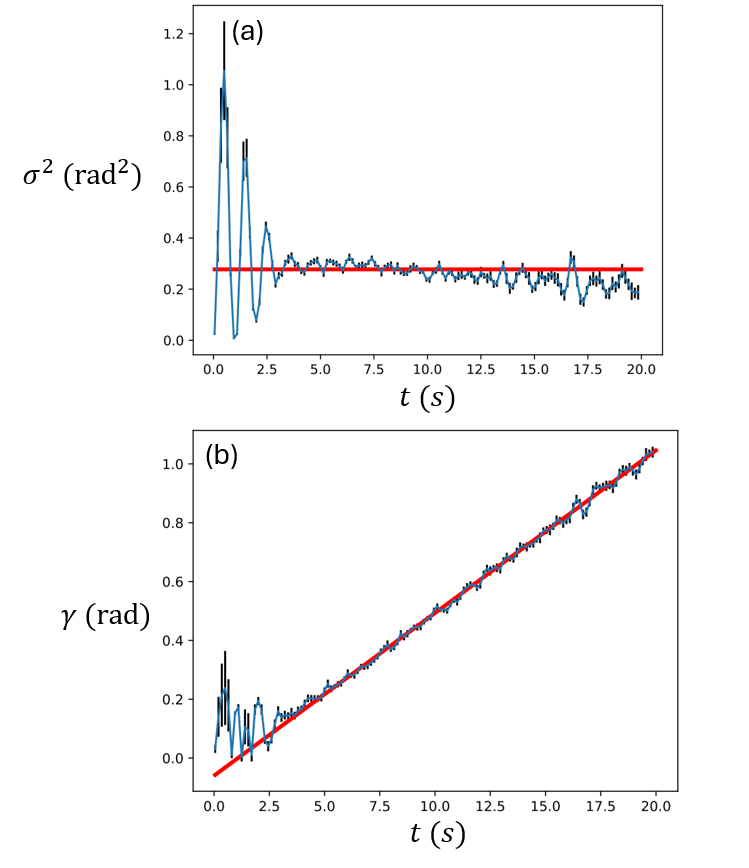}
    \caption{Voigt profile parameters as a function of time for Monte Carlo simulations of a helical ensemble with $D_G=0$, $D_L=0.05$, $\mu_A=0.5$, $\sigma_A=0.1$, $\mu_\omega=2\pi$, and $\sigma_\omega=0.75$. (a) $\sigma^2$ as a function of time with fit parameter in blue, error-bars on the fit in black, average value in red. There is more error early on since it does not follow a Voigt profile at early times. We see initial oscillations that damp out resulting in $\sigma^2$ becoming more-or-less constant. (b) $\gamma$ parameter as a function of time with fit parameter in blue, error-bars in black, and linear fit to $\gamma$ in red. Error is high early on but gets smaller with time. There are initial oscillations that damp out. At later times the behavior is linear. }
    \label{fig:Helical Voigt params}
\end{figure}

Figure~\ref{fig:HelicalTheory Example Distributions} shows this process for the parameter values given in Table~\ref{Table:Helical Params}. A movie of this distribution and Voigt profile fit can be seen in the supplementary material (S6). Early on the Voigt profile is not a good fit, the distribution splits and oscillates before eventually converging to a Voigt profile at long times (Fig.~\ref{fig:HelicalTheory Example Distributions}). Figure~\ref{fig:Helical Voigt params} shows the Voigt profile parameters as functions of time with parameter value in blue, error on the fit parameter (black bars), and either the average value (for $\sigma^2$) or a linear fit (for $\gamma$) in red. The Voigt profile parameter $\sigma^2$ (Fig.~\ref{fig:Helical Voigt params}a) shows initial oscillations that damp out and becomes more-or-less constant. The error in the oscillations is high early on, but becomes smaller with time, consistent with the distribution becoming more of a Voigt profile. The Voigt profile parameter $\gamma$  (Fig.~\ref{fig:Helical Voigt params}b) also shows some initial oscillations that damp out and become linear with time. The slope of this linear region is $0.0542\pm0.0001$, very close to the value of $D_L$ chosen in the simulations. 

\begin{table}[h]
\caption{Parameter values used in the Monte Carlo simulation for the helical ensemble.}
\label{Table:Helical Params}
\begin{tabular}{ || c | c | c | c | c | c ||}

\hline
$D_G$&$D_L$&$\mu_A$&$\sigma_A$&$\mu_\omega$&$\sigma_\omega$\\
\hline
$0.0$&$0.05$&$0.5$&$0.10$&$2\pi$&$0.75$\\
\hline
\end{tabular}
\end{table}

A more detailed analysis varying the parameters ($\mu_A$,$\sigma_A$) and ($\mu_\omega$,$\sigma_\omega$) while leaving everything else fixed shows interesting features in the fit parameters (see supplemental material for more detail). Varying the average amplitude $\mu_A$ controls the average value of $\sigma^2(t)$ in the Voigt profile fit. Varying the standard deviation $\sigma_A$ in amplitude across the ensemble controls the error in the fit during the initial oscillations, or how much it resembles a Voigt profile at early times. The average value in frequency $\mu_\omega$ controls the frequency of oscillations, with higher frequency at higher $\mu_\omega$. It also slightly shifts the average value of the Voigt parameter $\sigma^2(t)$. The standard deviation $\sigma_\omega$ in frequency across the ensemble controls the damping in the oscillations, with more damping for higher $\sigma_\omega$. It also plays a role in shifting the average value of the Voigt parameter $\sigma^2(t)$. This coupling between parameters that shifts the average value of $\sigma^2(t)$ makes it difficult to extract the means and standard deviations of the frequencies and amplitudes from the Voigt profile fit parameters. In all cases, however, the behavior of the Voigt profile parameter $\gamma(t)$ oscillates early on, damps out, and then grows linearly with slope very close to the value of $D_L$ chosen in the simulation.

Therefore, for helix-dominated  swimming behavior, such as what is observed in the \textit{Euglena} dataset, we expect initial oscillations that damp out resulting in parameter values that are more-or-less constant for the Voigt profile parameter $\sigma^2$ and linear for $\gamma$. We can measure the Lorentzian noise strength $D_L$ from the slope of the linear region in $\gamma$.

\section{\label{sec: Results} Time-dependence of fit parameters}

In this section we use the theory from Sec.~\ref{sec: Theory} to extract the noise strengths $D_L$ and $D_G$  of the various experimental microbial populations. This is achieved by fitting a Voigt profile at each time-step, and analyzing the parameters of the fit as a function of time. There are interesting qualitative features in the fits that are naturally explained by the application of the ensemble theory given in Sec.~\ref{Sec:Applications}.

\subsubsection{\label{subsec: wild params}Tumbling strain of \textit{Bacilus subtilis}}

We begin our estimation of the relevant parameter values for the microbial populations by analyzing the Voigt profile fits to the angular probability $P(\theta,t)$ for the tumbling strain of \textit{B. subtilis}  as a function of time. A video of the angular distribution (blue) with Voigt profile fit (red) can be seen in the supplementary material S7. It is striking how closely the Voigt profile fits the angular probability on the time-scales analyzed. 

\begin{figure}[h]
    \centering
    \includegraphics[width=1.0\columnwidth]{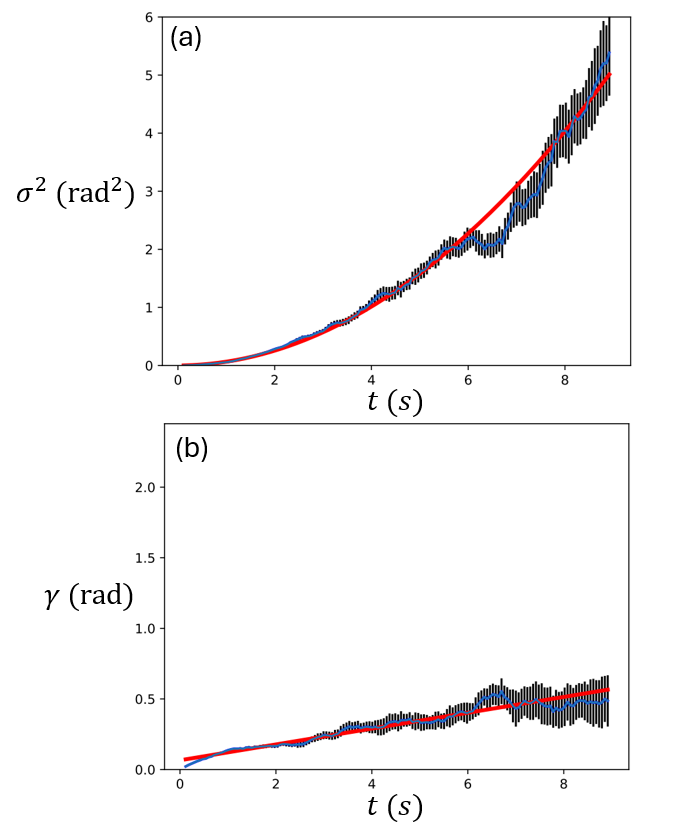}
    \caption{ Voigt profile parameters as a function of time for the expansion of the delta function of the angular probability for the tumbling strain of \textit{B. subtilis}. (a) $\sigma^2$ as a function of time with fit parameter in blue, error-bars on the fit in black, and quadratic fit to $\sigma^2$ in red. There is clear quadratic behavior in $\sigma^2$, which is expected from Gaussian distributed ensembles of circular arcs. (b) $\gamma$ parameter as a function of time with fit parameter in blue, error-bars in black, and linear fit to $\gamma$ in red. It expands rapidly and then becomes linear. This behavior is expected from ensembles with helical trajectories. The Gaussian part dominates the Voigt profile at longer times.}
    \label{fig:WildType VoigtParams}
\end{figure}

Figure~\ref{fig:WildType VoigtParams} shows the behavior of $\sigma^2$ and $\gamma$ as functions of time (blue) with error estimates for the fit (black). The parameter $\sigma^2$ (Fig.~\ref{fig:WildType VoigtParams}a) shows quadratic behavior in time.  We fit a quadratic polynomial to this plot (red). Recall that \textit{B. subtilis} moves in circular arcs. Thus, using the theory from Sec.~\ref{Sec:Applications} we expect $\sigma^2$ to be the polynomial
\begin{align}
\sigma^2=\sigma_\omega^2t^2+2D_G t.
\end{align}
The parameter $\sigma^2$ roughly follows this behavior, with a very small constant term. Thus we estimate both the Gaussian noise strength ($D_G$), and the standard deviation in the ballistic angular velocity term ($\sigma_\omega$) across the ensemble in Table~\ref{Table:Wild-type Noise Params}. We can estimate the order of magnitude for the expected Gaussian noise strength from the physical rotational diffusion for a spherical particle given by the Stokes-Einstein relation
\begin{align}
D_{\text{rot}}=\frac{k_B T}{8 \pi \eta R^3},\label{Stokes-Einstein}
\end{align}
where $k_B$ is the Boltzmann constant, $T$ is the temperature of the fluid, $\eta$ is the dynamic viscosity of the fluid, and $R$ is the radius of the particle. The fluid is at room temperature ($T=293-295$ K). The fluid used is Lysogeny broth (LB) medium, with dynamic viscosity $\eta\approx10^{-3}$ Pa$\cdot$s, and the typical length of \textit{B. subtilis} is $2-6$~$\mu$m. We therefore estimate $D_{\text{rot}}=O(10^{-3})-O(10^{-1})$ rad$^2$/s, which means $D_G$ (Table~\ref{Table:Wild-type Noise Params}) is the correct order of magnitude. 

Figure~\ref{fig:WildType VoigtParams}b shows the $\gamma$ parameter as a function of time. Roughly linear behavior is observed, with a faster initial expansion, which is most likely caused by the presence of some helical trajectories. From the theory in Sec.~\ref{sec: SDE/FPKE} we can estimate the Lorentzian noise strength given by 
\begin{align}
    \gamma=D_L t.
\end{align}
We fit a line to the linear region after the initial expansion (red). Table~\ref{Table:Wild-type Noise Params} shows the extracted parameter value (the slope of the line). 

\begin{table}[h]
\caption{Extracted parameters for the tumbling strain of \textit{B. subtilis}\label{Table:Wild-type Noise Params}}
\begin{tabular}{ ||c | c | c ||}

\hline
$D_G$ (rad$^2$/s)& $D_L$ (rad/s)& $\sigma_\omega$ (rad/s)\\
\hline
$0.0012\pm0.0003$& $0.056\pm0.001$&$0.250\pm0.001$ \\  
\hline  
\end{tabular}
\end{table}

Since $D_L<\sigma_\omega$, the Voigt profile for the tumbling \emph{B. subtilis} is always Gaussian dominated. The timescale where the Gaussian part of the Voigt profile switches from being dominated by noise, to being dominated by the circular arcs is $t_B\approx 0.04$ s. Thus, for virtually all times, the distribution of ballistic angles dominates the statistics.

\subsubsection{\label{subsec: Smooth Bac}Smooth swimming strain of \textit{Bacilus subtilis}}

Next we estimate the relevant parameter values for the smooth swimming strain of \textit{B. subtilis}, which tends to swim more smoothly than the tumbling strain. We fit a Voigt profile to the angular probability $P(\theta,t)$ and analyze the behavior of $\sigma^2$ and $\gamma$ as functions of time. A video of the angular distribution (blue) with Voigt profile fit (red) can be seen in the supplementary material S8. It is again very striking how well the Voigt profile fits the data over time. 

\begin{figure}[h]
    \centering
    \includegraphics[width=1.0\columnwidth]{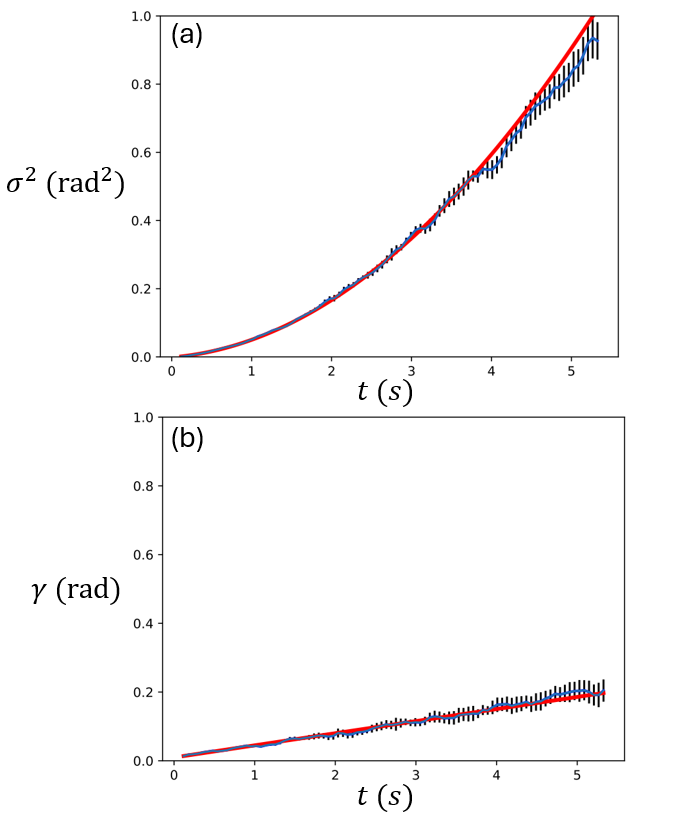}
    \caption{ Voigt profile parameters as a function of time for the expansion of the delta function for the angular probability for the smooth swimming strain of \textit{B. subtilis} (a) $\sigma^2$ as a function of time with fit parameter in blue, error-bars in black, and quadratic fit to $\sigma^2$ in red. There is clear quadratic behavior in $\sigma^2$, this is again expected from Gaussian distributed ensembles of circular arcs. (b) $\gamma$ parameter as a function of time with fit parameter in blue, error-bars in black, and linear fit to $\gamma$ in red.There is no fast initial expansion in $\gamma$ for this population. The Gaussian part dominates the Voigt profile at longer times.}
    \label{fig:SmoothBac VoigtParams}
\end{figure}

Figure~\ref{fig:SmoothBac VoigtParams} shows the behavior of $\sigma^2$ and $\gamma$ as functions of time (blue) with error estimates for the fit (black). Similar to the tumbling strain of \textit{B. subtilis} the parameter $\sigma^2$ (Fig.~\ref{fig:SmoothBac VoigtParams}a) shows quadratic behavior in time. We therefore fit a quadratic polynomial to this plot (red). Similar to the tumbling strain, this population of microbes moves in circular arcs. Thus using a similar argument, we can estimate the Gaussian noise term ($D_G$) and the standard deviation in the ballistic angular velocity term ($\sigma_\omega$) over the ensemble shown in Table~\ref{Table:SmoothBac Noise Params}. The fluid used in the experiments for this population is CAP medium~\cite{Rummens85} with dynamic viscosity close to that of water ($\eta\approx10^{-3}$ Pa$\cdot$s). The temperature and type of fluid is roughly the same as the tumbling strain experiments. We estimate $D_{\text{rot}}=O(10^{-3})-O(10^{-1})$ rad$^2$/s using the Stokes-Einstein relation given by Eq.~(\ref{Stokes-Einstein}). Thus $D_G$ again has the correct order of magnitude. 

Next we estimate the strength of Lorentzian noise. Figure~\ref{fig:SmoothBac VoigtParams}b shows that we get linear behavior in $\gamma$. This plot does not show a fast initial expansion. This may mean distributions of helices are less prevalent in this ensemble. Thus we estimate $D_L$ in Table~\ref{Table:SmoothBac Noise Params}.

\begin{table}[h]
\caption{Extracted parameters for the smooth swimming strain of \textit{B. subtilis}\label{Table:SmoothBac Noise Params}}
\begin{tabular}{ ||c | c | c ||}

\hline
$D_G$ (rad$^2$/s)& $D_L$ (rad/s)& $\sigma_\omega$ (rad/s)\\
\hline
$0.0084\pm0.0002$& $0.035\pm0.005$&$0.184\pm0.001$ \\ 
\hline
\end{tabular}
\end{table}

Since $D_L<\sigma_\omega$, the Voigt profile is again, always Gaussian dominated. The timescale where the Gaussian part of the Voigt profile switches from being dominated by noise to being dominated by the circular arcs is $t_B\approx 0.5$ s. Thus, we can observe the region in time where the statistics are dominated by Gaussian noise. Comparing to the parameter values of the tumbling strain, we observe that both $D_L$ and $\sigma_\omega$ are smaller, which is expected if this population swims straighter and more smoothly. The Gaussian noise term is larger than the Gaussian noise term for the tumbling strain.  This could be attributed to differences in stochasticity in the swimming behavior (rather than thermal rotational diffusion) or differences in the sizes of the microbes between strains.

\begin{figure}[h]
    \centering
    \includegraphics[width=1.0\columnwidth]{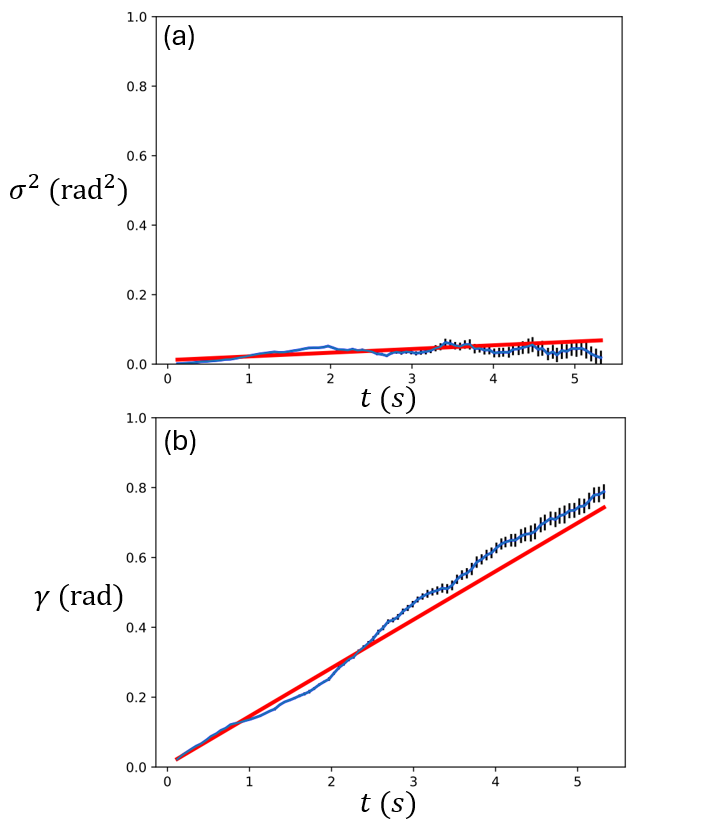}
    \caption{ Voigt profile parameters as a function of time for the expansion of the delta function for the angular probability for \textit{Tetraselmis} (a) $\sigma^2$ as a function of time with fit parameter in blue, error-bars in black, and linear fit to $\sigma^2$ in red. We roughly linear behavior in $\sigma^2$ this is  expected from Gaussian noise most likely caused by physical rotational diffusion. (b) $\gamma$ parameter as a function of time with fit parameter in blue, error-bars in black, and linear fit to $\gamma$ in red. The Lorentzian part dominates the Voigt profile at longer times.}
    \label{fig:Tetra VoigtParams}
\end{figure}

\subsubsection{\label{subsec: Tetraselmis}\textit{Tetraselmis suecica}}

Here, we estimate the relevant parameter values for \textit{Tetraselmis}.  We fit a Voigt profile to the angular probability $P(\theta,t)$ and analyze the behavior of $\sigma^2$ and $\gamma$ as functions of time. A video of the angular distribution (blue) with Voigt profile fit (red) can be seen in the supplementary material S9. The Voigt profile is a strikingly good fit. However, at later times, the Voigt profile becomes less good of a fit to the tails. This may be attributed to errors in measurements, smaller sample-sizes as time progresses, or the model we offer is only a good approximation to the angular statistics on the timescales we analyzed.

Figure~\ref{fig:Tetra VoigtParams} shows the behavior of $\sigma^2$ and $\gamma$ as functions of time (blue) with error estimates for the fit (black). Figure~\ref{fig:Tetra VoigtParams}a shows the behavior of $\sigma^2$ as a function of time. The plot has roughly linear behavior with a faster linear region from $t=0$ s to $t=2.0$ s, and a flatter region after $t=2.0$ s. From classical physical diffusion 
\begin{align}
\sigma^2=2D_G t.
\end{align}
We estimate the Gaussian noise strength by a weighted least squares linear fit to the whole time-period. Table~\ref{Table:Tetra Noise Params} shows the estimated Gaussian noise strength ($D_G$). The experiments were done at room temperature ($T=293-295$ K) in Alga-Gro sea water medium with dynamic viscosity close to water ($\eta\approx10^{-3}$ Pa$\cdot$s). The typical size of \textit{Tetraselmis} is $10-15$ $\mu$m. We therefore estimate the order of magnitude for the physical rotational diffusion as $D_{\text{rot}}=O(10^{-4})-O(10^{-3})$ rad$^2$/s, Eq.~(\ref{Stokes-Einstein}). Thus, $D_G$ has the correct order of magnitude.

Next we estimate the Lorentzian noise strenth $D_L$. Figure~\ref{fig:Tetra VoigtParams}b shows the behavior of $\gamma$ as a function of time. There is  roughly linear behavior in $\gamma(t)$ with some oscillations, most likely due to the presence of some helices. We fit a line (red) to the data and estimate $D_L$ in Table~\ref{Table:Tetra Noise Params}.

\begin{table}[h]
\caption{Extracted parameters for \textit{Tetraselmis}\label{Table:Tetra Noise Params}}
\begin{tabular}{ ||c | c ||}

\hline
$D_G$ (rad$^2$/s)& $D_L$ (rad/s)\\
\hline
$0.0053 \pm 0.0005 $& $0.138\pm0.002$ \\  
\hline  
\end{tabular}
\end{table}

There is a timescale in this system where the Voigt profile switches from being Gaussian dominated to Lorentzian dominated. For this microbe it is $t_L\approx0.6$~s. We can observe this early on in the distributions, where it is more Gaussian at early times, but as time goes on it becomes more Lorentzian. Comparing to the other microbes $D_G$ is on the same order as both \textit{B. subtilis} populations, this behavior is unexpected since the size of the microbe is larger. If the only Gaussian noise in this system is physical diffusion we expect on average a lower Gaussian noise strength. Therefore, we speculate that the relatively large Gaussian noise strength, which is the same order as the bacteria, could be related to the activity rather than physical Brownian motion due to the fluid properties. The strength of Lorentzian noise is much larger, however, which is expected since \textit{Tetraselmis} tumbles much more dramatically. 

\begin{figure}[h]
    \centering
    \includegraphics[width=1.0\columnwidth]{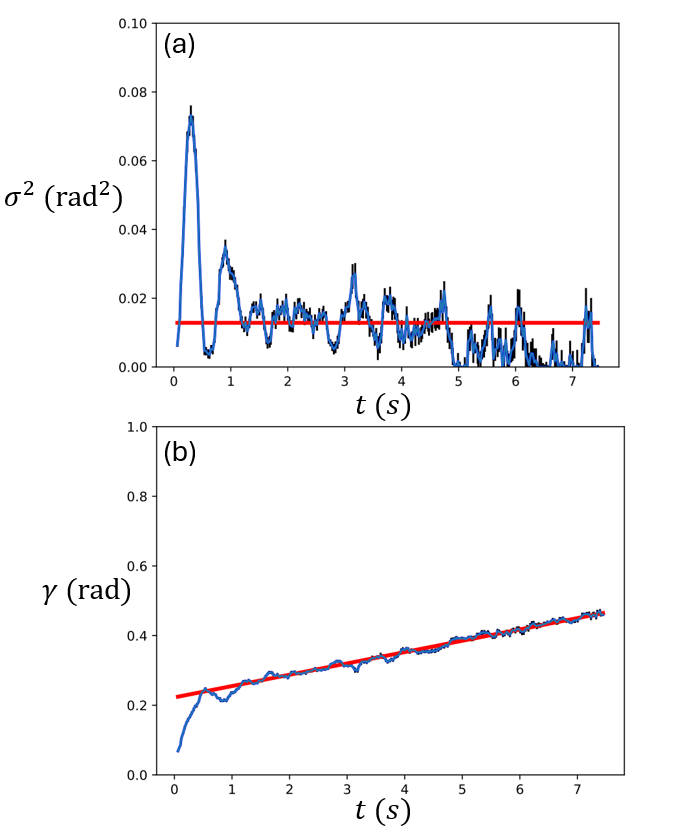}
    \caption{ Voigt profile parameters as a function of time for the expansion of the delta function for the angular probability for \textit{Euglena} (a) $\sigma^2$ as a function of time with fit parameter in blue, error-bars in black, and average value $\sigma^2$ in red. The initial oscillations may be due to the helical natures of the trajectories. The oscillatory behavior in $\sigma^2$  damps out and becomes noisy. There is not clear growth in $\sigma^2$ overtime. This may be due to the microbe generally not moving in circular arcs, and having a negligible physical rotational diffusion constant due to the typical size of \textit{Euglena} being relatively large. (b) $\gamma$ parameter as a function of time with fit parameter in blue, error-bars in black, and linear fit to the linear region of $\gamma$ in red. There is a fast initial expansion in $\gamma$, that then bounces, and is followed by linear behavior. This may be due to the helical nature of the trajectories. The Lorentzian part of the Voigt profile dominates the distribution.}
    \label{fig:Euglena VoigtParams}
\end{figure}

\subsubsection{\label{subsec: Euglena}\textit{Euglena gracilis}}

Finally we estimate the relevant parameter values for \textit{Euglena}. We fit a Voigt profile to the angular probability $P(\theta,t)$ and analyze the behavior of $\sigma^2$ and $\gamma$ as functions of time. A video of the angular distribution (blue) with Voigt profile fit (red) can be seen in the supplementary material S10. The Voigt profile is a strikingly good fit to the data. 

Figure~\ref{fig:Euglena VoigtParams} shows the behavior of $\sigma^2$ and $\gamma$ as functions of time (blue) with error estimates for the fit (black). Figure~\ref{fig:Euglena VoigtParams}a shows the behavior of $\sigma^2$ as a function of time. There is oscillatory behavior early on that damps out and becomes noisy and small. Recall that this oscillatory behavior and damping is expected from ensembles of helices with varying frequency, amplitude, and phase. The experiments are done at room temperature ($T=293-295$ K). The fluid used for \textit{Euglena} is soil-water medium with dynamic viscosity close to that of water ($\eta\approx10^{-3}$ Pa$\cdot$s). Euglena is relatively large compared to the other microbes at a length of $40-80$ $\mu$m. Thus, using the Stokes-Einstein relation, we estimate $D_{\text{rot}}=O(10^{-6})-O(10^{-5})$~rad$^2$/s, which is probably negligible, leading to the relatively constant behavior in $\sigma^2$ after the initial oscillations.

We extract the strength of Lorentzian noise ($D_L$) for this population of microbes. Figure~\ref{fig:Euglena VoigtParams}b shows the behavior of $\gamma$ as a function of time. There is  a fast initial expansion that then oscillates and grows linearly. Recall that for ensembles of helices this oscillatory behavior is expected. The parameters for the distributions of helices set where the linear region will begin. Recall that the slope of the linear region after the fast initial expansion is very close to the parameter value $D_L$ in the Monte Carlo simulations. Thus we estimate $D_L$ in Table~\ref{Table:Euglena Noise Params}. 
\begin{table}[h]
\caption{Extracted Lorentzian noise parameter for \textit{Euglena}\label{Table:Euglena Noise Params}}
\begin{tabular}{ ||c ||}

\hline
$D_L$ (rad/s)\\
\hline
$0.0325\pm0.0003$ \\  
\hline  
\end{tabular}
\end{table}
Since the expected Gaussian noise is so small, and the amplitude of oscillations in $\sigma^2$ is small, we expect this population to almost always be Lorentzian dominated. We do not observe a deviation early on from the Voigt profile, thus we expect the standard deviation in amplitude of the helices $\sigma_A$ to be relatively large compared to the average amplitude $\mu_A$.  Comparing the value of $D_L$ to the other populations, we observe that it is comparable to the smooth swimming strain of \textit{B. subtilis}, and is smaller than that of the tumbling strain of \textit{B. subtilis} and \textit{Tetraselmis}.

\section{\label{sec: Conclusion}Conclusion}

We have shown that we can use a Lorentzian noise model to describe tumbling dynamics. That is, micro-swimmers undergo L\'{e}vy walks in orientation with L\'{e}vy index $\alpha=1$. This description of tumbling is ubiquitous for several microbes including bacteria and marine algae. We may speculate on why the Lorentzian noise model is so robust in describing tumbling dynamics. The Lorentzian distribution often appears when describing resonant phenomena, where the heavy tails of the distribution represent large fluctuations about the resonant value (the center of the distribution). A notable physical example includes spectral lines in spectroscopy. Resonance of some form, either resonance related to the activity (e.g. proton pump), or resonance of the motion of the flagella may lead to this distribution being observed in the experimental data. 

Theoretically the tumbling dynamics for smaller microbes is coupled with additive Gaussian noise most likely due to either physical rotational diffusion or additional noise (other than Lorentzian) related to the swimming behavior, which leads to a Voigt profile for the orientation statistics.  Note, however, that the Gaussian part of the Voigt profile is not always dominated by Gaussian noise.  Instead, it can sometimes be dominated by the variations in deterministic behavior in the ensembles. Each swimmer has different deterministic behavior, such as circular arcs with different sizes or helices with different amplitudes, frequencies, and phases, and the statistical nature of this varying behavior averaged over the ensemble leads to measurable effects on the Voigt profile parameters as a function of time.

\section{\label{sec: Acknowledgements}Acknowledgments}

 This work was supported by NSF grant DMR-2302708, NSF grant CMMI-2314417, as well as funding through the UC Merced Center for Cellular and Biomolecular Machines (CCBM) NSF-CREST (NSF grant HRD-2112675).

\end{document}